\documentclass[conference]{IEEEtran}
\usepackage[utf8]{inputenc}
\usepackage[pdftex]{graphicx}
\usepackage{subfigure}
\usepackage{graphicx}
\usepackage{amsmath}
\usepackage{tabularx, booktabs}
\usepackage{amssymb}
\usepackage{makecell}
\usepackage{algorithm}
\usepackage{algpseudocode}
\usepackage{graphics}
\usepackage[dvips]{epsfig}
\usepackage{listings}
\usepackage{xspace}
\usepackage{multirow}
\usepackage{mathtools}
\usepackage{amsmath}
\usepackage[dvipsnames]{xcolor}

\usepackage{soul}
\usepackage{url}

\usepackage{xcolor}
\definecolor{bluedark}{RGB}{0,0,127}
\definecolor{greendark}{RGB}{0,100,0}
\definecolor{brown}{RGB}{165,42,42}
\definecolor{bluemedium}{RGB}{0,0,205}
\definecolor{gray}{RGB}{128, 128, 128}
\definecolor{darkorchid}{RGB}{153,50,204}

\begin{document}
\title{A Software-Defined Networking approach for congestion control in Opportunistic Networking}

\author{\IEEEauthorblockN{1\textsuperscript{st}MªCarmen de Toro} 
\IEEEauthorblockA{\textit{Dept. of Information and Communications Engineering} \\ 
CYBERCAT-Center for Cybersecurity Research of Catalonia\\
Universitat Autònoma de Barcelona. Barcelona, Spain.\\ 
\textit{mariacarmen.detoro@uab.cat}}
\and
\IEEEauthorblockN{2\textsuperscript{nd}Carlos Borrego}
\IEEEauthorblockA{\textit{Dept. of Information and Communications Engineering} \\
CYBERCAT-Center for Cybersecurity Research of Catalonia\\
Universitat Autònoma de Barcelona. Barcelona, Spain.\\ 
\textit{carlos.borrego@uab.cat}}

}

\maketitle
\begin{abstract}
The short-term adoption of opportunistic networks (OppNet) depends on improving the current performance of this type of network. Software-Defined Networks (SDN) architecture is used by Internet applications with high resource demand. SDN technology improves network performance by programmatically managing the network configuration by using a control layer.
In this paper, we propose that OppNet nodes use a control layer to get an overview of the whole network and use this knowledge to apply policies to get a better performance of the network. As a use case for our experimentation, we have focused on improving congestion control in OppNet with a control layer that dynamically regulates the replication degree used by forwarding algorithms. We have compared the performance of our proposal with two different configurations of the OppNet, over two community scenarios based on real mobility traces. The results of the test prove that our SDN-like approach overruns the other two approaches in terms of delivery ratio and latency time performance.
\end{abstract}


\section{Introduction}

Opportunistic Networking (OppNet) \cite{mota2014protocols} is a consolidated network paradigm where there is no guarantee of an end-to-end path between nodes in the network. In this context, data flows from origin to destination using \textit{ad hoc} opportunistic contacts between nodes.
This network paradigm has been proved to be a good solution for a wide number of scenarios, like sensor networks, the Internet of Things, unmanned aerial vehicle networks and interplanetary networks.  It has also been applied in networks where their users need to maintain the privacy of their whereabouts or their social contacts \cite{borrego2019efficient, borrego2019hey,  borrego2017softwarecast}. Although OppNet is a communication solution where other networks fail to provide a network service, there are still some issues to be addressed before they can be widely deployed.

OppNet uses a scheme of store carry and forward (SCF) for data forwarding. Most OppNet forwarding algorithms use replication strategies to achieve proper delivery ratios and low latency times. Both principles have proved to be the most effective way to achieve good network performance \cite{mota2014protocols}. Conversely, replication could turn to be a double-edge strategy causing congestion and affecting the performance of the network. 

In the scope of network architectures, Software Defined Networks, (SDN) \cite{kreutz2014software} is a paradigm used to increase network performance in high-throughput scenarios. This paradigm uses a component named controller that is aware of the use of the entire network as well of the resources to be managed. The controller (control layer), gathers network information pro-actively by running network application services on the underlying virtual or physical switches of the network (infrastructure layer). Therefore, with this information, the controller dynamically assigns network resources to the applications.

Our hypothesis in this study is to prove that using an approach of the SDN control layer over an OppNet, will yield network performance improvement. Hence, for achieving the network awareness, we propose nodes in the OppNet acting as controllers in the SDN sense. The rest of the nodes in the network act pro-actively by gathering and sending network information to the controllers. These nodes take measurements about network performance indicators, the forwarding algorithms they use, and the setup parameters of those forwarding algorithms. By so doing, the nodes perform what we call \textit{network crowdsensing}. There is a direct analogy between these nodes and the sensors in a wireless sensor network (WSN). Sensors in a WSN periodically send their collected data to a sink. Thereby the SDN-like OppNet can be seen as a WSN where the sensors are the network nodes acting both as sensors and routers \cite{biswas2004opportunistic}.  

In this work, we present a strategy, based on the combination of the use of a SDN-like control layer with network crowdsensing. We test the feasibility and performance of our approach in the scope of the congestion control issue.


\section{Related Work}

In this section, we study the state of the art of the Opportunistic Networks (OppNet). We focus on forwarding and congestion control. We introduce the Software Defined Networks (SDN) architecture.

\subsection{Opportunistic Networks}

An opportunistic network, OppNet, \cite{mota2014protocols} is a network where nodes have high mobility, where the contact time between nodes is unpredictable and where the end-to-end path between nodes is not guaranteed. Under these conditions, the TCP/IP protocol suite used in the Internet cannot be applied. 
There is still active research on finding techniques, algorithms and services to improve network performance in such a challenging environment. 

In the following subsections, we focus on two main research topics in OppNet, forwarding strategies and congestion control.

\subsubsection{Traffic forwarding in OppNet}

In OppNet, traditional link layer connection-oriented solutions for traffic forwarding are not suitable as these networks are topology-agnostic. Instead, OppNet uses the store-carry and forward paradigm (SCF)  for hop to hop message relay. The SCF forwarding algorithms commonly use the replication mechanism (multiple-copy forwarding) as a forwarding strategy to improve delivery rates and decrease end-to-end delivery latency time \cite{vahdat2000epidemic, spyropoulos2005spray, borrego2015mobile}. 

In \cite{mota2014protocols}, the authors present a study of the current OppNet forwarding algorithms. They conclude that network's context awareness forwarding algorithms, perform better than the ones which do not use any context information. 
In OppNet, there are two different levels of context: the link level and the social level. At link level, we can be aware of what is happening physically in the network regarding congestion, overhead, delivery ratio and latency. The social context is built as a result of gathering information of the node contacts.  

\subsubsection{Congestion control in OppNet}
In OppNet, nodes have limited resources: limited buffer size, limited energy consumption and limited contact duration. 

There are two types of congestion in a communication network: the link congestion, and the node storage or buffer congestion.
In an opportunistic network, the link congestion is rare due to the nodes' mobility. Thus, it is just considered storage congestion.

In OppNet, a congestion scenario is likely to happen depending on the SCF forwarding algorithm applied by the nodes \cite{soelistijanto2013transfer}. Most of the SCF forwarding algorithms are based on message replication, which implies multiple copies of the message in the network. Replicating messages increase the network throughput and, therefore, the limited resources of the nodes can be overwhelmed, resulting in the node's buffer over-flow, which decreases the delivery rates. In this case, the nodes drop incoming or already stored messages depending on their policies. Existing message dropping strategies use per-node local available data as message priority, message lifetime, message size, or message delivery probability. Each node needs to choose carefully the message drop strategy to be used to achieve a good network delivery ratio \cite{pan2013comprehensive}. 

\subsection{Software Defined Networking}

Many circumstances necessarily force changes in the present-day Internet, such as the fast increase in the use of cloud computing, the extensive use of multimedia streaming or the high-speed, high-throughput and low-latency Internet of Things applications requirements. IP networks are based on predefined built policies and are very difficult to configure to be dynamically adaptive to changes.  Software-Defined Networking (SDN) \cite{kreutz2014software} is a new architecture that offers flexibility, scalability and adaptability for high-throughput internet applications at a very effective cost, by using simplified hardware, software, and management.
The fundamental principle of the SDN paradigm is that applications' or services' network requirements define the optimal network settings in terms of routing, bandwidth, lifetime, priorities and policies. 

Current IP networks are vertically integrated as the control and data planes are bundled together. Control plane manages setting up routing flows, making routing decisions and applying network policies. Once the control plane has configured data flows, they are pushed down to the data plane. Data plane handles data forwarding and data processing from the underlying routers and switches at a physical level. 
SDN architecture entirely separates the data and the control planes in a horizontal integration way. Promotes the use of a  software, referred to as controller, which has the global view of the network. Thereby, the applications, through the controller, see the network as a single logical switch. 
The controller is a program, or programs, running on a server, that use a vendor-neutral well-defined API as OpenFlow \cite{openFlow} to gather information used by the control plane to define high-level flow rules to be applied by the data plane.  The controller behaves as a single element despite its possible distributed implementation.
The controller programs the network i.e. sets up the control plane dynamically based on the network state and the application's network requirements. Setting up the control plane consists of programming packet-processing rules matching multiple header fields from layer 2 to layer 4,  and performing multiple actions.

Nevertheless, decoupling the control and data planes adds overhead to the communications as control information is generated between the controller and data plane (southbound communication). In return, programming the controller makes network management easier and more flexible than configuring the network at the interface level. With this decoupling between data and control plane, changes in the network's underlying infrastructure have a low impact on the applications.  

In \cite{li2017mobile}, Li et al., apply the SDN paradigm in OppNet in what they call Software Defined Opportunistic Networks (SDON). They implement a SDON to build a mobile crowdsensing system. 
In a SDON, the data plane consists of the connections intra-mobile devices and between mobile devices and access points. However, SDN architecture has been conceived to be applied to a wired infrastructure to ensure reliable communication between the data plane and the controller. To guarantee southbound communication reliability, Li et al. use a cellular network to send the control messages while data is sent opportunistically. This hybrid approach has some privacy drawbacks. By using a cellular network, the infrastructure provider knows the position of the node, and thus the owner of the node.   

As we have seen so far, in \cite{mota2014protocols}, the authors state that context-based forwarding algorithms perform better than non-context-based ones. SDN use context information by using a controller and have proved to perform well over scenarios of high network throughput. Li et al. in \cite{li2017mobile} implement a SDON where southbound communication is done via a connected network, which implies some security drawbacks. With those premises, we propose a SDN-like architecture entirely opportunistic. We name this approach controlled OppNet. We evaluate how it performs in the context of congestion control. 

\section{Controlled OppNet Architecture}


From the SDN architecture, we have taken on the concept of the control layer, where controllers, implemented as network services, manage to have an overview of the status of the network. With this knowledge, the controllers send to the nodes in the OppNet the actions to be done to improve the performance of the network. The controllers generate those actions based on their current perception of the state of the network.

in OppNet, all the nodes in the network opportunistically forward messages whenever they have contact with other nodes. The decision of which nodes perform as controllers depends on the nature of the network. For instance, in a vehicular ad hoc network, the controller could be running in a roadside unit (RSU). In a pocket switch network, the controller could emerge from an influencer node. Nonetheless, the main idea is that any node in the OppNet can run the control service.

Adopted from the concept of crowdsensing, we consider that all the nodes in the OppNet act as network sensors in terms of retrieving network performance indicators. They periodically send this information to the controllers.  The controllers themselves are part of the OppNet, and therefore also perform this network sensing. 

We consider two types of control messages for the communication between controllers and nodes:
\begin{itemize}
    \item \textbf{Network Metric:} Network-specific indicators sensed by the nodes are encapsulated in a network metric message and sent to the controllers.
    \item \textbf{Control Directive:} The controllers use the metric messages received from the nodes to have an overview of the network. With this information, if the controllers decide that an actuation has to be done to improve the performance of the current state of the network, the controllers send a directive message encapsulating network configuration parameters. When the nodes receive those directives, they apply, to the new messages they create, the network configuration encapsulated in the received directive.
\end{itemize}    

\begin{figure}[!htp]
    \includegraphics[width=1.0\columnwidth]{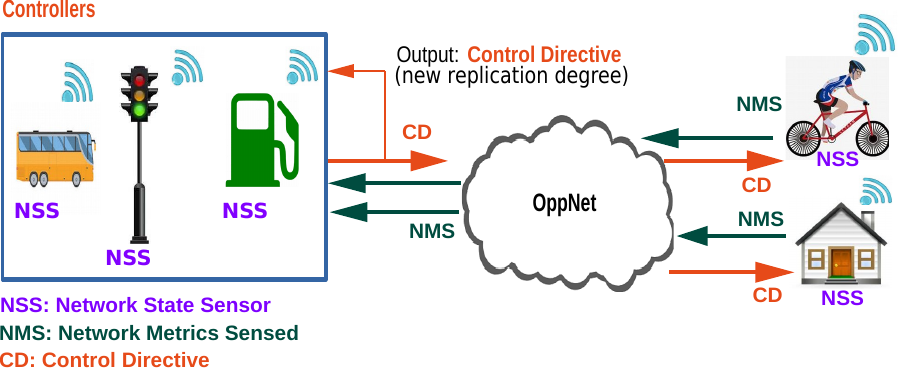}
    \caption{Controllers receive the network metrics sensed by the nodes. With this information, the controllers send a directive to the nodes to adjust their replication degree. The controllers apply themselves this directive.}
    \label{fig:control_general}
\end{figure}

Fig. \ref{fig:control_general} shows our concept of a controlled OppNet and the messages exchanged between nodes and controllers.

\subsection{Congestion control using a controlled OppNet}
The use case we consider to test the controlled OppNet is the congestion control.

As we have mentioned in the introduction section, the forwarding algorithms used by OppNet, mostly use the principle of replication as a technique to get a better delivery performance. When there is a contact between nodes, their implemented forwarding algorithms decide under which degree of replication they forward the messages to the contacted nodes. This replication degree determines the maximum number of nodes that have a copy of the message. This replication can go from an epidemic one, meaning the message is forwarded to any contact that isn't carrying the message already, to more restrictive techniques, where just a limited amount of copies of the message are forwarded throughout the network. \cite{poonguzharselvi2013survey}.

We propose that all nodes, including the controllers, sense the number of dropped messages during a window time  (wt\_messageDropsReading). After this window time, the nodes generate a metric encapsulating the number of registered dropped messages (dropped messages metric). They opportunistically send to the controllers this metric. The controllers gather those metrics for a window time (wt\_metricsReading). After this window time, with all the received metrics, the controllers infer the degree of congestion of the network. Depending on this degree, the controllers send a directive to all nodes specifying a new value for the message replication degree to be used by the forwarding algorithm. If the controller detects that there is no congestion in the network, it sends a directive to the nodes to increase the value of the degree of replication. By doing so, when the nodes create new messages, more copies of the messages are allowed in the network, and the messages have more chances to get to the destination. On the contrary, if the controller detects that there is congestion in the network, it sends a directive to the nodes to decrease the value of the replication degree. This way, when the nodes create new messages, fewer copies of the messages are allowed in the network, and the congestion level decreases.

\subsection{Control System Logic}

\begin{figure}[!ht]
    \centering
    \includegraphics[width=0.8\columnwidth]{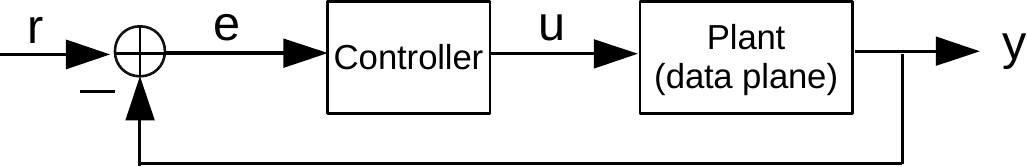}
    \caption{Feedback closed loop control system. Where \textbf{r:} \textit{reference signal}  = no congestion; \textbf{e:} \textit{control error} = r – congestion; \textbf{u:} \textit{manipulated control variable} = new replication degree; \textbf{y:} \textit{process controlled variable} = congestion sensed.}
    \label{ourFeedback}
\end{figure}

We have implemented the control service as a closed-loop feedback control system \cite{dorf2011modern} as shown in Fig. \ref{ourFeedback}. 
A closed-loop feedback control system is a control system that maintains a constant relation between the output of the system (\textit{y}: the controlled variable) and the desired output (\textit{r}: the reference signal), by subtracting one to the other as a measure of control.

Our controller is a proportional one (P-controller) \cite{janert2013feedback} with the following control function: 

\vspace*{-0.35cm}
\begin{equation*}
 f(RD\textsubscript{t}, congestion) = RD\textsubscript{t+1} 
\end{equation*}
\vspace*{-0.35cm}
where
\vspace*{-0.35cm}
\begin{equation*}
    RD\textsubscript{t+1} = RD\textsubscript{t} - K \times congestion \times (-1)^\delta
\end{equation*}    
\begin{equation*}    
    \begin{cases}
        \delta = 1 & \text{if congestion} < \text{threshold}\\
        \delta = 2 & \text{if congestion} \geq \text{threshold}
    \end{cases}
\end{equation*}
\noindent
where \textit{RD\textsubscript{t+1}} is the Replication degree at time t+1 and \textit{K} is the proportional factor of the control action. This value is specified as a configuration parameter. The \textit{congestion} parameter is calculated aggregating all the received metrics from the nodes for a window time. To aggregate these metrics it is used an exponentially weighted moving average function (EWMA) as follows. For all received metrics in a window time:

\vspace*{-0.35cm}
\begin{equation*}
congestion = (1-\alpha) \times congestion + \alpha \times congestion\_metric  
\end{equation*}
where \textit{alpha} is a configuration setting specifying the proportional relation between the weight we give to the already aggregated \textit{congestion} and the weight we give to the new reading of congestion (\textit{congestion\_metric}).

\begin{algorithm}[t!]
\scriptsize
\caption{Control service logic.}
\label{alg:control}
\begin{algorithmic}[1]
\Procedure{send\_directive\_to\_nodes}{}
\If{RD\_current not set}
 \State RD\_current = RD\_default
\EndIf
\State congestion = null
\State RD\_from\_other\_ctrls\_avg = null
\While{window\_time}
\State congestion\_metric = recv(from\_nodes)
\State congestion = EWMA(congestion, congestion\_metric)   
\State directive\_from\_other\_ctrls = recv(from\_ctrls)
\State RD\_from\_other\_ctrls\_avg=EWMA(RD\_from\_other\_ctrls\_avg, directive\_from\_other\_ctrls)

\EndWhile
\If{congestion $>=$ threshold}
 \State RD\_new = DECREASE(RD\_current)
\Else
 \State RD\_new = INCREASE(RD\_current)    
\EndIf 
\State RD\_new = EWMA(RD\_new, RD\_from\_other\_ctrls\_avg)
\State send(RD\_new, to\_nodes)
\EndProcedure

\Function {EWMA}{value, valueToBeAggregated}
  \State\Return $(1-\alpha) * value + \alpha * valueToBeAggregated$
\EndFunction
\Function{decrease}{value}
  \State new\_value = value * K 
\State \Return{new\_value}
\EndFunction
\Function{increase}{value}
  \State new\_value += value * K 
\State \Return{new\_value}
\EndFunction

\end{algorithmic}
\end{algorithm}

Algorithm \ref{alg:control} shows the flow of the logic of the control service. Each control cycle of the controller (lines 1 to 20 in Algorithm \ref{alg:control}), lasts a window time. The output of the control system after this control cycle is the new replication degree (\textit{RD\_new}) to be sent, as a directive, to the nodes. Once a node receives a directive, applies the new replication degree to the new messages the node creates, either data, directive or metric messages.

During a window time, the controller receives the congestion metric sent by the nodes (lines 7 to 12 in Algorithm \ref{alg:control}). These received metrics are aggregated through an EWMA. The directives the controller receives from other controllers during this window time are aggregated using another EWMA (lines 10 to 11 in Algorithm \ref{alg:control}). These directives encapsulate the replication degree calculated by other controllers. At the end of the window time, the resulted aggregated congestion value is used to decide whether to increase or decrease the configured replication degree (lines 13 to 17 in Algorithm \ref{alg:control}). Once the replication degree has been adjusted, it is aggregated through another EWMA to the already-mentioned aggregated directives being received from other controllers (line 18 in Algorithm \ref{alg:control}). The final calculated replication degree is encapsulated in a directive, and it is sent to the nodes (line 19 in Algorithm \ref{alg:control}).

\section{Experimentation}

\begin{center}
\begin{figure}[t]
\begin{center}
  \includegraphics[width=0.5\textwidth]{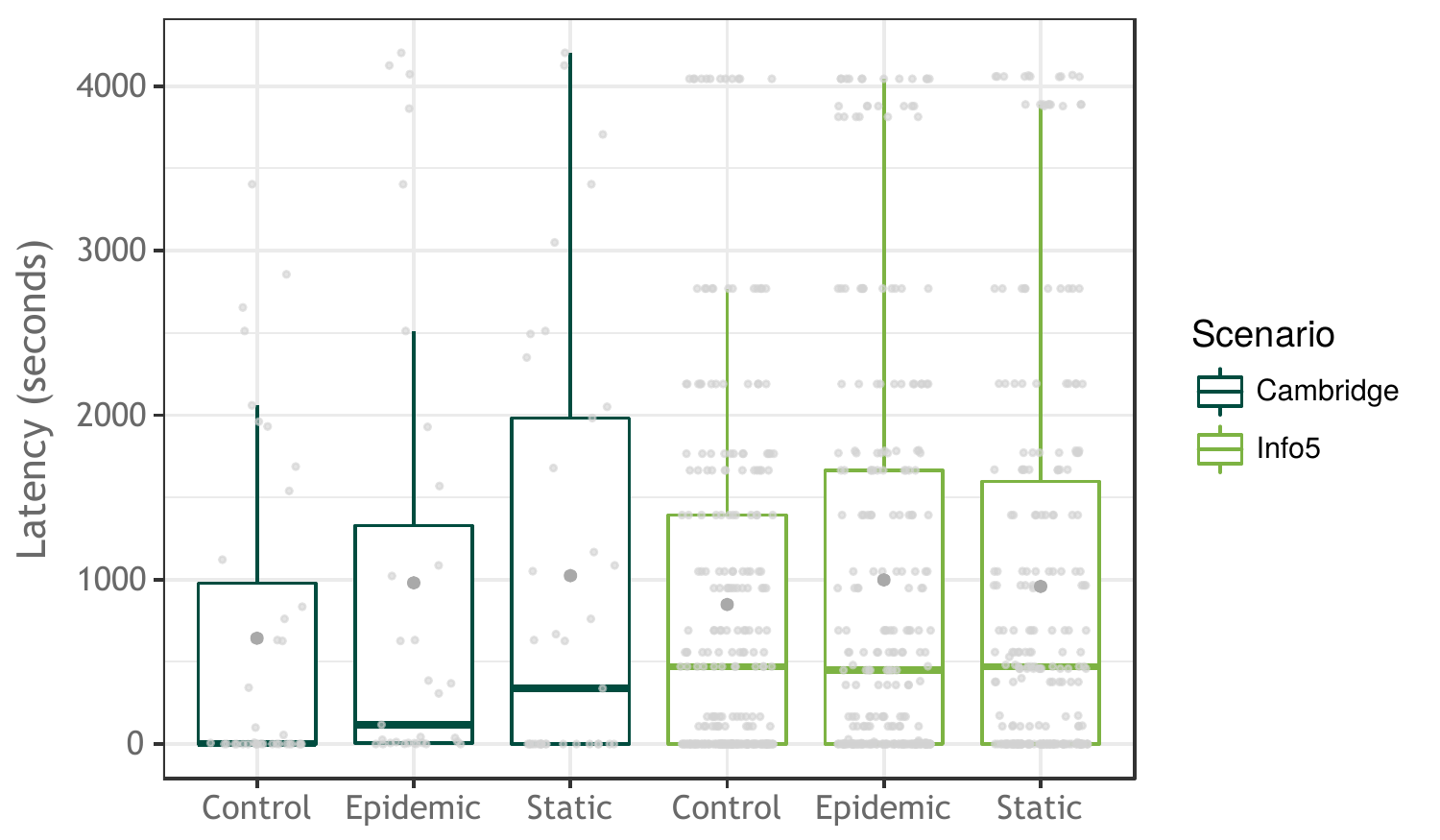}
  \caption{Latency time. The y-axis measures the messages' latency time  distribution for the Control, Epidemic, and Static configurations for Cambridge and Info5 scenarios (x-axis).}
  \label{fig:latency}
\end{center}
\end{figure}
\end{center}

\begin{center}
\begin{figure}[t]
\begin{center}
  \includegraphics[width=0.5\textwidth]{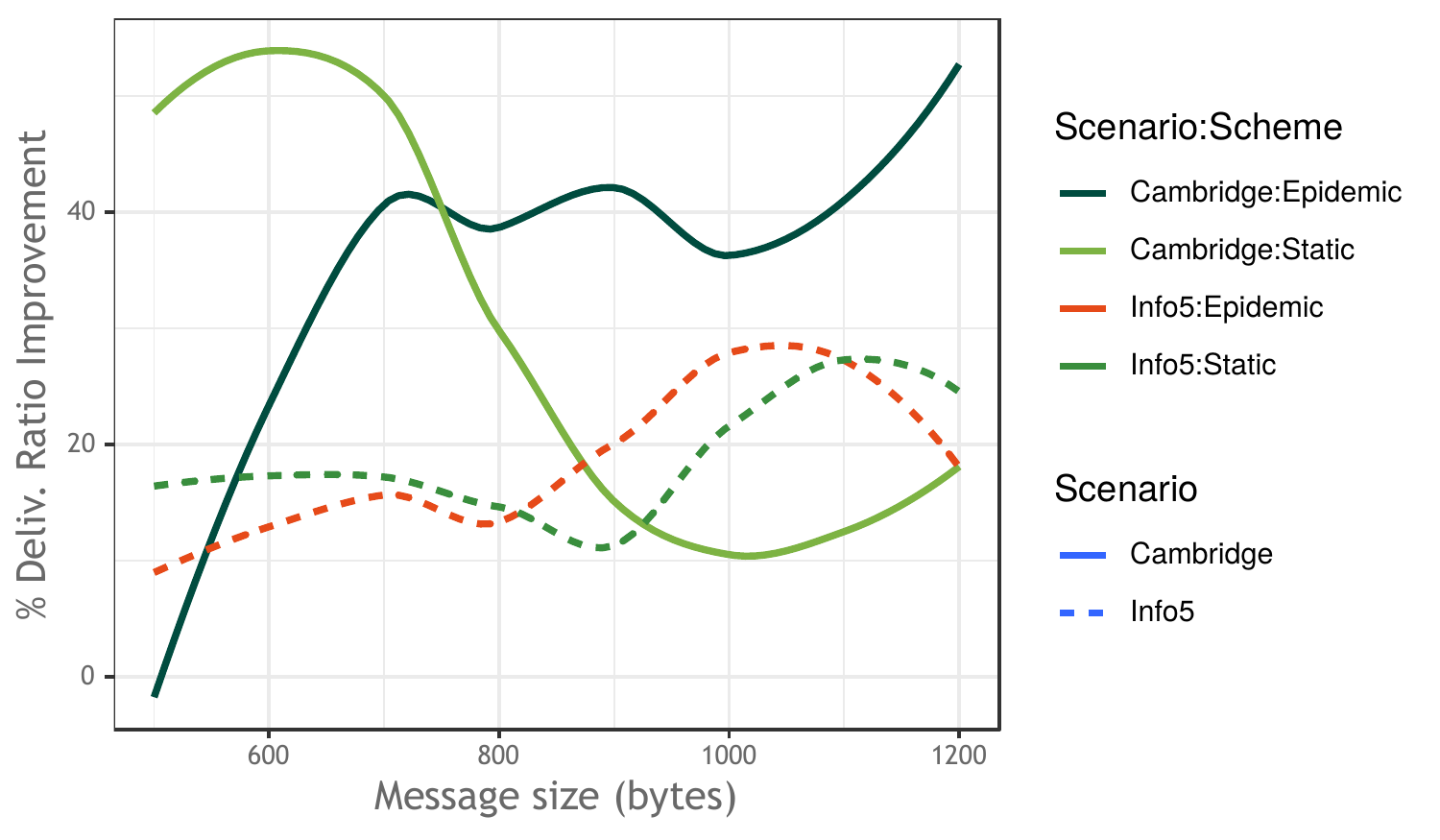}
  \caption{Delivery ratio improvement. The y-axis measures the delivery ratio improvement of the Controlled OppNet over the Epidemic and the Static approaches for different message sizes (x-axis).}
  \label{fig:delivery}
\end{center}
\end{figure}
\end{center}

In the following subsections, we introduce which are the network performance indicators we consider. We describe the network scenarios used for the simulation and the configuration settings for those scenarios. Finally, we present the achieved results.

\subsection{Performance indicators}
The simulation indicators we consider to measure the performance of the network are:
\begin{itemize}
    \item \textbf{Message delivery ratio:} It measures the percentage of the created messages that get to the destination.
    \item \textbf{Message latency:} It measures the time it has taken a message to get to its destination.
\end{itemize}

\subsection{Simulation scenarios and configuration}
The experimentation has been performed by extending\footnote{The source code of our TheONE extension is available at \cite{theOne_control}.} the Opportunistic Network Environment (TheONE) simulator \cite{TheOne} with a control layer.

Our controlled OppNet uses a forwarding algorithm based on replication. The replication degree used by the forwarding algorithm changes dynamically depending on the saturation level of the buffers of the nodes in the network. We compare the performance of our controlled OppNet, we refer as Control, with the same OppNet configuration, without the control layer, using two different forwarding strategies. The first OppNet no-controlled configuration uses a replication-based forwarding algorithm applying a static configured replication-limit of 10 copies of the message. We refer to this configuration as Static. The second configuration uses an epidemic forwarding algorithm. We refer to this configuration as Epidemic. 

We have considered two scenarios to run the performance tests. Both scenarios define the node contacts obtained from real mobility traces from the Crawdad database\footnote{\textit{http://crawdad.org/keyword-DTN.html}.}, a community resource for collecting wireless data at Dartmouth College, United States. The first scenario we have used is the \textit{Cambridge} scenario. It is a community scenario formed by real contact traces from 51 students from the System Research Group of the University of Cambridge in the UK. The students carried small devices during six days \cite{upmc-content-20061117}. Also, static nodes were located at different points of interest as the student's computer Lab, the local commerce, cultural and entertainment locations around the city. The second scenario we have used is the \textit{Info5} scenario. It is a community scenario based on real mobility traces from 41 students attending the Infocom conference in 2005. The traces \cite{uoi-haggle-20160828} were gathered during 2.93 days of the conference. 

TABLE \ref{tab:simulation} shows the basic settings used for the simulation.

\begin{table}[th!]
\caption{Simulation settings summary.}
  \centering
\begin{tabularx}{0.48\textwidth}{l l l}
\toprule
\textbf{Common Settings} & \textbf{Value} \\
\midrule
Transmission Speed & 100MB/s & \\
Transmission Range & 100m & \\
Buffer Size & 30MB & \\
Data Message Size & 600B To 1MB & \\
Data Message Generation Interval & 25s To 35s & \\
\midrule
\textbf{Specific Control Settings} & \textbf{Value} \\
\midrule
Metric Generation Interval &  60s\\
Directive Generation Interval &  90s\\
EWMA Alpha for a New Metric Reading & 0.8 & \\
Proportional Control Factor (K) & 0.2 & \\
Metric Message Size & 21B & \\
Directive Message Size & 5B & \\
Congestion Threshold & 10 drops & \\
\midrule
\textbf{Specific Scenario Settings} & \textbf{Cambridge} & \textbf{Info5} \\
\midrule
Simulation Time & 8h & 48h \\
\# Nodes & 52 & 41 \\
\# Contacts & 1378 & 18786 \\
\bottomrule
\end{tabularx}
\label{tab:simulation}
\end{table}

\subsection{Simulation Results}
Firstly, we evaluate the performance of our controlled OppNet compared to the Static and Epidemic ones in terms of delivery ratio. Finally, we evaluate the performance in terms of latency time. 

\subsubsection{Delivery ratio performance}
Fig. \ref{fig:delivery} shows the percentage of the delivery ratio improvement of the Control compared to the Static and the Epidemic ones for Cambridge and Info5 scenarios (y-axis), for different data message sizes (x-axis). 
For the Cambridge scenario, comparing the results between the Control and the Epidemic configurations, we appreciate that for small-sized data messages, the Control's delivery ratio is close to the Epidemic one. However, when the size of the messages starts to increase, the delivery ratio of the Control tends to overcome 50\% of improvement compared to the Epidemic. The heavier the messages are, the more significant is the network throughput, and an epidemic replication strategy ends up with the nodes buffers' overflow. In this situation, our adaptive controlled approach regulates the replication degree, which reduces network throughput and, therefore, gets a better delivery ratio. Comparing the results of the Control and the Static configurations, we appreciate that for small-sized messages, the Control approach performs up to 50\% better than the Static one. This result is due to the Static's conservative replication degree. As the size of the messages rises, the network throughput increases, and there is more congestion in the nodes. In this situation, the Static approach improves its delivery ratio, as a few copies of the messages in the network are allowed, which helps in reducing congestion. Nevertheless, the Control approach still performs better than the Static one, although in a less percentage compared to lighter messages. Those results entail that the Control configuration would perform better with a more strict replication strategy in case of congestion for this scenario.

For the Info5 scenario, for all message sizes, the Control approach outperforms the Epidemic one. Again, those results relay on the congestion of the network resulting from the epidemic replication behaviour. Compared to the Static approach, the Control one improves its delivery ratio for all the message sizes. Specifically, the best improvement occurs when there is congestion in the network. The Control configuration has been able to adapt the replication degree according to the current network congestion, and therefore, has outperformed the Static configuration.

As we have seen, the dynamic controlled replication strategy adapts to the different scenarios, always getting a delivery ratio improvement compared to the Static and the Epidemic approaches. 

\subsubsection{Latency time performance}
Fig. \ref{fig:latency}, in the y-axis, displays a box plot showing the distribution of the message latency time in the range of the first quartile (Q1=25\%) and the third one (Q3=75\%), and the minimum and maximum values for the Control, Epidemic and Static configurations for Cambridge and Info5 scenarios (x-axis). 

Regarding the Cambridge scenario, the messages' latency time for the Control approach is shorter than the Epidemic one. As we see from the delivery ratio improvement chart, the Epidemic approach performance suffers from congestion, which implies higher latency times for the messages. The replication strategy of the Static approach helps with the delivery ratio but does not do so good with messages' latency time, as there are fewer copies of the message in the network. In the case of the Info5 scenario, the congestion caused by the Epidemic directly affects its latency time. The limit of copies of the Static is not suitable to get a good latency time.

For both scenarios, the Control approach gets the best performance in terms of latency time, as regulates the number of message copies, reducing them in case of congestion, but also increasing them when there is no congestion ahead. 

The traffic overhead generated by the control layer is minimal for both scenarios. The measured control traffic overhead for Cambridge is $3.9\text{e-}{4}$, and for the Info5 one is $3.8\text{e-}{4}$. 

\section{Conclusions}
\label{sec:conclusions}
We have developed a controller service to run over OppNet nodes. The controller service receives, for a window time, the network congestion metrics sent by the nodes. After this window time, the controller service, with these metrics, calculates the new optimal replication level according to the congestion measured. The controller service sends, as a directive, the new replication level to the nodes in the network. The nodes that receive the directive apply it to the new messages they create. 
We have tested three different configurations over the Cambridge and info5 community scenarios. The first configuration uses the control service. The second configuration uses an epidemic forwarding protocol, and the third configuration uses a forwarding protocol based on a static replication degree. The simulations results prove that our controlled approach, with a dynamic adaptive replication system, performs better in terms of improving the delivery ratio of data messages and decreasing its latency time compared to the Static and the Epidemic configurations. The control service achieves such improvement with a minor traffic overhead.

\section*{Acknowledgment}

This work is partly funded by the Catalan AGAUR 2017SGR-463 project, and by the Spanish Ministry of Science and Innovation TIN2017-87211-R project.


\bibliography{dtn}{}
\bibliographystyle{IEEEtran}
\end{document}